\documentclass[twocolumn,showpacs,aps,prd]{revtex4-1}
\usepackage{graphicx}
\usepackage{dcolumn}
\usepackage{amsmath}
\usepackage{epsfig}
\usepackage{subfigure}
\usepackage{epstopdf}
\usepackage{multirow}
\usepackage{mathrsfs}
\usepackage{colortbl}
\usepackage[bookmarks=false] {hyperref}
\usepackage{cleveref}
\usepackage{lineno}
\usepackage{ulem}
\usepackage{enumerate}
\newcommand{\piz}{\pi^{0}}

\newcommand{\psipp}{\psi(3770)}

\newcommand{\ks}{K_{S}^{0}}

\newcommand{\dhhee}{D\rightarrow K \pi e^{+}e^{+}}

\newcommand{\dztokpill}{D^{0} \rightarrow K^{-} \pi^{-} l^{+} l^{+}}
\newcommand{\dptokzpill}{D^{+} \rightarrow \bar{K}^{0}\pi^{-} l^{+} l^{+}}
\newcommand{\dztokpilldcs}{D^{0} \rightarrow K^{-}\pi^{-} l^{+} l^{+}}
\newcommand{\dptokpizll}{D^{+} \rightarrow K^{-}\pi^{0} l^{+} l^{+}}
\newcommand{\dkpipipi}{D^0\rightarrow K^{-} \pi^{+} \pi^{+} \pi^{-}}
\newcommand{\dkpipizero}{D^0 \rightarrow K^{-} \pi^{+} \pi^{0}}
\newcommand{\dkpipi}{D^+ \rightarrow K^{+} \pi^{+} \pi^{-}}

\newcommand{\eeqqbar}{e^{+}e^{-} \rightarrow q\bar{q}}

\begin{document}


\title{\boldmath Search for a heavy Majorana neutrino in lepton number violating decays of $D\to K \pi e^+ e^+$}
\date{\today}

\author{
\begin{small}
\begin{center}
M.~Ablikim$^{1}$, M.~N.~Achasov$^{10,d}$, S. ~Ahmed$^{15}$, M.~Albrecht$^{4}$, M.~Alekseev$^{56A,56C}$, A.~Amoroso$^{56A,56C}$, F.~F.~An$^{1}$, Q.~An$^{53,43}$, J.~Z.~Bai$^{1}$, Y.~Bai$^{42}$
% ...
\\
\vspace{0.2cm}
(BESIII Collaboration)\\
\vspace{0.2cm} {\it
$^{1}$ Institute of High Energy Physics, Beijing 100049, People's Republic of China\\
% ...
$^{4}$ Bochum Ruhr-University, D-44780 Bochum, Germany\\
% ...
$^{10}$ G.I. Budker Institute of Nuclear Physics SB RAS (BINP), Novosibirsk 630090, Russia\\
% ...
$^{15}$ Helmholtz Institute Mainz, Johann-Joachim-Becher-Weg 45, D-55099 Mainz, Germany\\
% ...
$^{42}$ Southeast University, Nanjing 211100, People's Republic of China\\
$^{43}$ State Key Laboratory of Particle Detection and Electronics, Beijing 100049, Hefei 230026, People's Republic of China\\
% ...
$^{53}$ University of Science and Technology of China, Hefei 230026, People's Republic of China\\
% ...
$^{56}$ (A)University of Turin, I-10125, Turin, Italy; (B)University of Eastern Piedmont, I-15121, Alessandria, Italy; (C)INFN, I-10125, Turin, Italy\\
% ...
$^{d}$ Also at the Novosibirsk State University, Novosibirsk, 630090, Russia\\
% ...
}\end{center}

\vspace{0.4cm}
\end{small}
}

\begin{abstract}
Using a data sample with an integrated luminosity of 2.93 fb$^{-1}$ taken at the center-of-mass energy of 3.773 GeV, we search for the Majorana neutrino ($\nu_m$) in the lepton number violating decays $D\to K \pi e^+ e^+$. No significant signal is observed, and the upper limits on the branching fraction at the 90\% confidence level are set to be $\mathcal{B}\,(D^0 \to K^- \pi^- e^+ e^+)<2.8\times10^{-6}$, $\mathcal{B}\,(D^+ \to K_S^0 \pi^- e^+ e^+)<3.3\times10^{-6}$ and $\mathcal{B}\,(D^+ \to K^- \pi^0 e^+ e^+)<8.5\times10^{-6}$. The Majorana neutrino is searched for with different mass assumptions ranging from 0.25 to 1.0 GeV/$c^2$ in the decays $D^0\to K^- e^+ \nu_m,~\nu_m \to \pi^- e^+$ and $D^+\to K_S^0 e^+ \nu_m,~\nu_m \to \pi^- e^+$, and the upper limits on the branching fraction at the 90\% confidence level are at the level of $10^{-7} \sim 10^{-6}$, depending on the mass of the Majorana neutrino. The constraints on the mixing matrix element $|V_{e\nu_m}|^2$ are also evaluated.
\end{abstract}


\maketitle
\section{\boldmath INTRODUCTION}
In the Standard Model (SM), due to the absence of a right-handed neutrino component and the requirements of SU(2)$_\mathrm{L}$ gauge invariance and renormalizability, neutrinos are postulated to be massless. However, the observations of neutrino oscillation~\cite{osci1,osci2,osci3,theta13} have shown that neutrinos have a tiny mass, which provides the first evidence for physics beyond the SM. Theoretically, the leading model to accommodate the neutrino masses is the so-called ``see-saw" mechanism, which can be realized in several different schemes~\cite{seesaw1,seesaw2,seesaw3,seesaw}. In the canonical case, the mass ($m_{\nu}$) of an observed light neutrino is given by $m_{\nu} \sim y^{2}_{\nu} \upsilon^{2}/m_{\nu_{m}}$, where $y_{\nu}$ is a Yukawa coupling of a light neutrino to the Higgs field, $\upsilon$ is the Higgs vacuum expectation value in the SM, and $m_{\nu_{m}}$ is the mass of a new massive neutrino state $\nu_m$. The smallness of $m_{\nu}$ can be attributed to the existence of the new neutrino state $\nu_m$ with high mass.

The nature of neutrinos, whether neutrinos are Dirac or Majorana particles, is still an open question. If they are Majorana, their particles and antiparticles are identical, while they are not identical if they are Dirac particles.
The effects of Majorana neutrino can be manifested through the processes violating lepton-number ($L$) conservation by two units ($\Delta L =2$). Consequently, experimental searches for Majorana neutrinos occurring through lepton-number violating (LNV) $\Delta L=2$ processes are of great interest. Different $\Delta L =2$ processes at low and high energies have been proposed in the literatures~\cite{feynman1,glopez,donghairong,lihb_rare_baryon,diego}. Among them, an interesting source of LNV processes is given by exchanging a single Majorana neutrino with a mass on the order of the heavy flavor mass scale, where the Majorana neutrino can be kinematically accessible and produced on shell. The effects of such a heavy neutrino with mass in the range 100 MeV/$c^{2}$ to a few GeV/$c^{2}$ have been widely searched for in $\Delta L$ = 2 three-body and four-body decays of heavy flavor mesons and in $\tau$ lepton decays by different experiments, as summarized in Ref.~\cite{uppersummary}, but no evidence has been observed so far. The $\Delta L$ = 2 processes of $D$ mesons have been reported by the E791 collaboration~\cite{E791} with upper limits (ULs) on the decay branching fraction (BF) ranging $10^{-5} \sim 10^{-4}$.

\begin{figure*}[htbp]
  \centering
   \subfigure[$\mathrm{~\dztokpill}$~(CF)]{\label{fig:subfig:a}\includegraphics[width=0.22\textwidth]{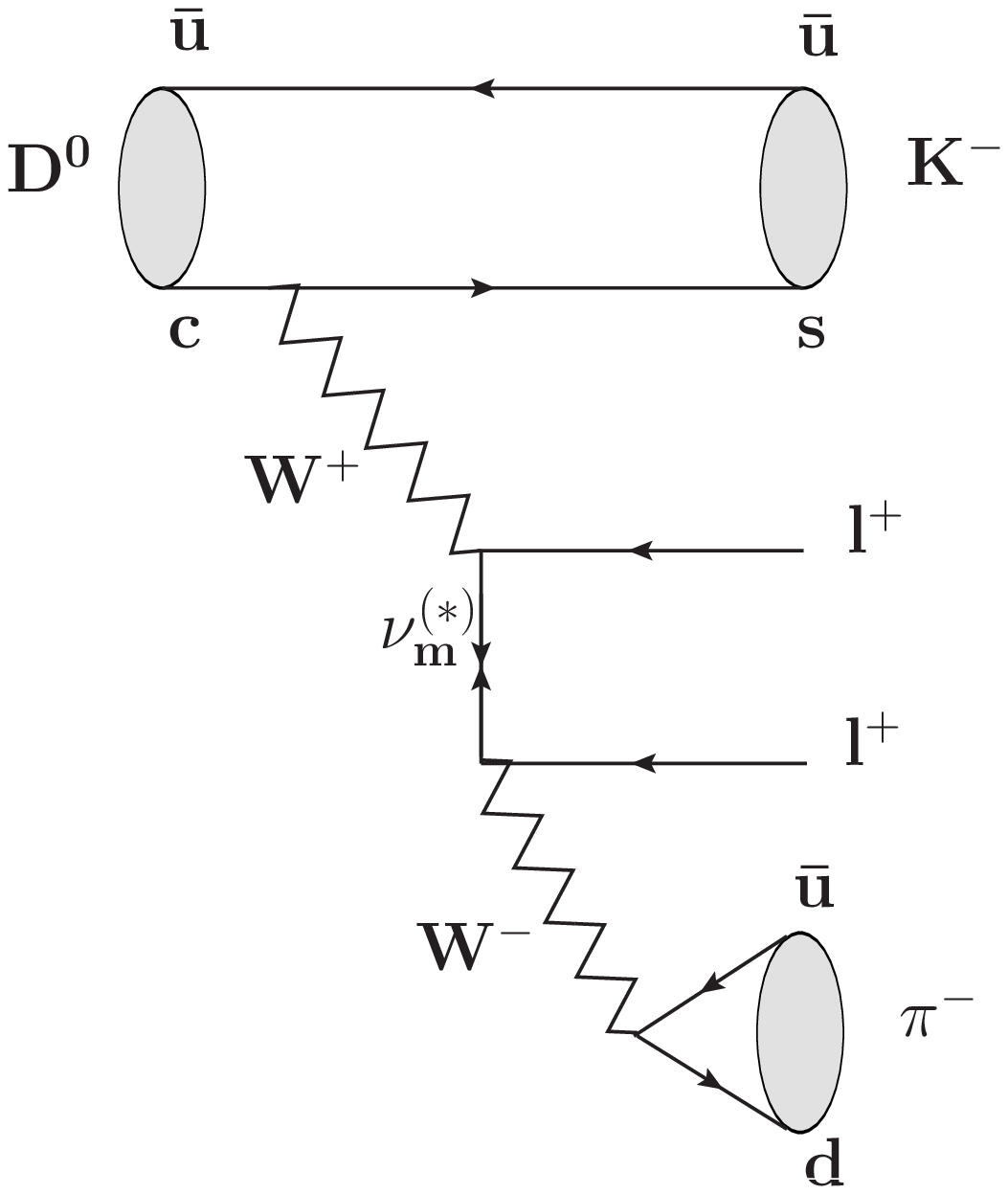}}
   \subfigure[$\mathrm{~\dptokzpill}$~(CF)]{\label{fig:subfig:b}\includegraphics[width=0.22\textwidth]{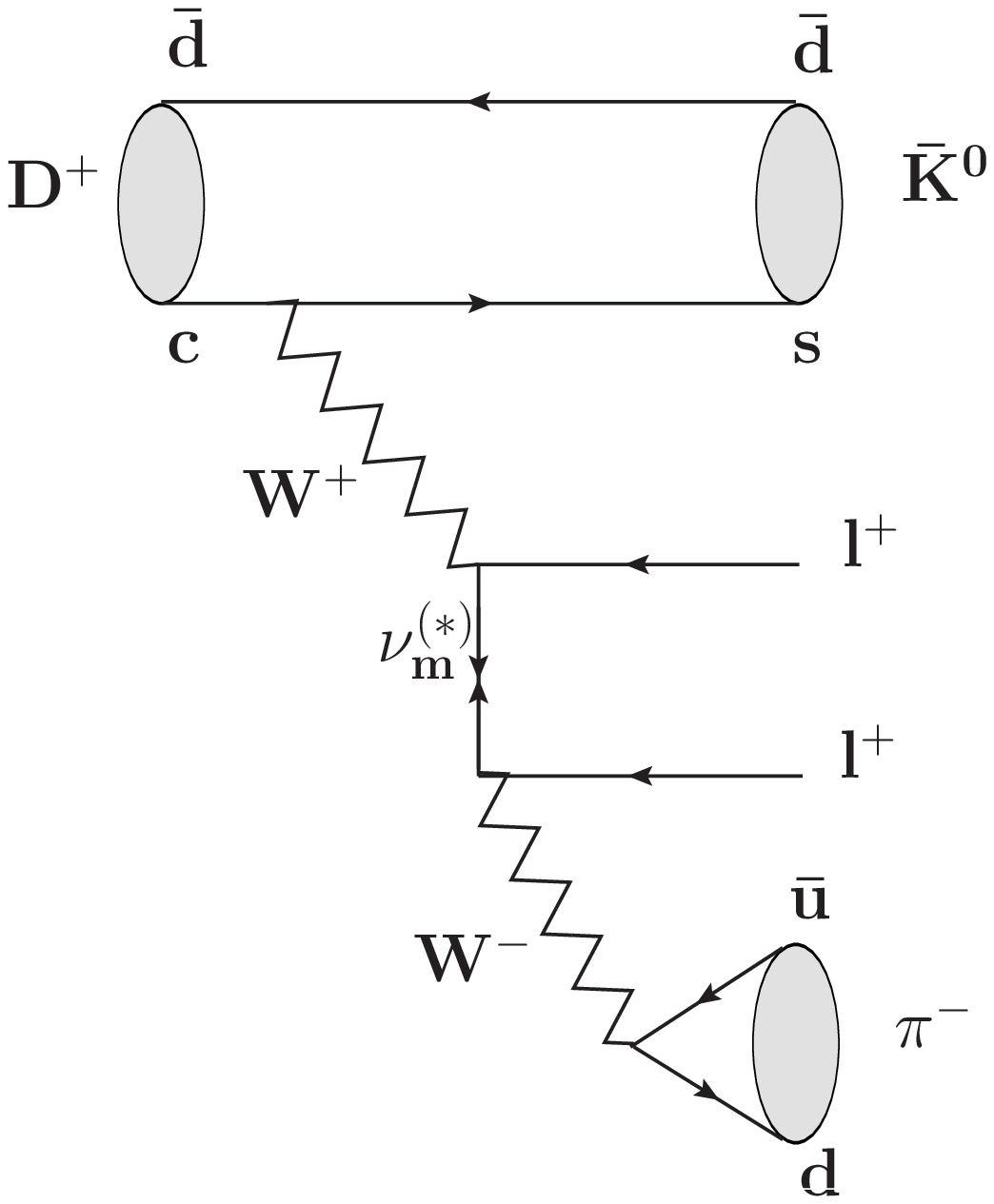}}
   \subfigure[$\mathrm{~\dztokpilldcs}$~(DCS)]{\label{fig:subfig:c}\includegraphics[width=0.22\textwidth]{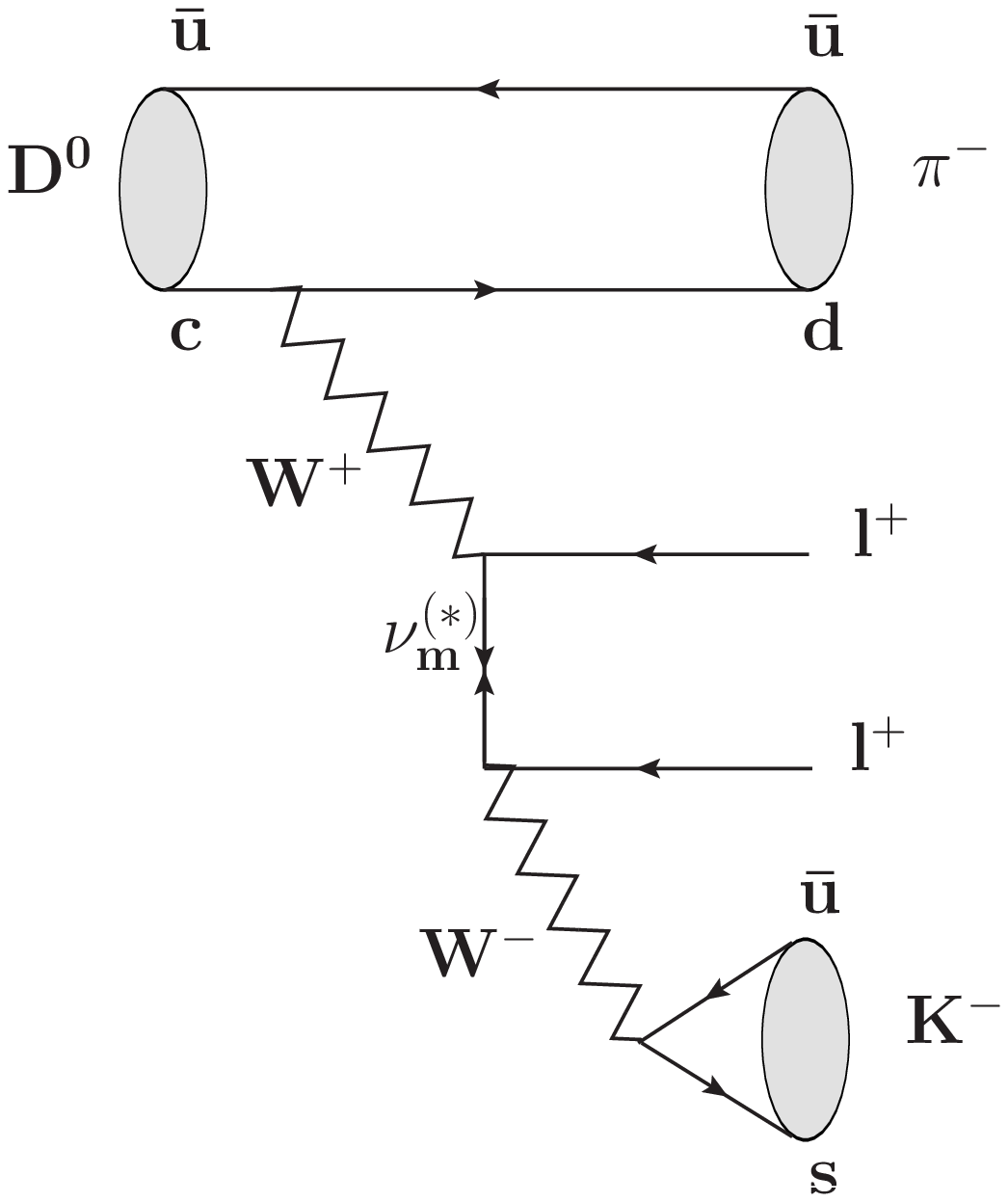}}
   \subfigure[$\mathrm{~\dptokpizll}$~(DCS)]{\label{fig:subfig:d}\includegraphics[width=0.22\textwidth]{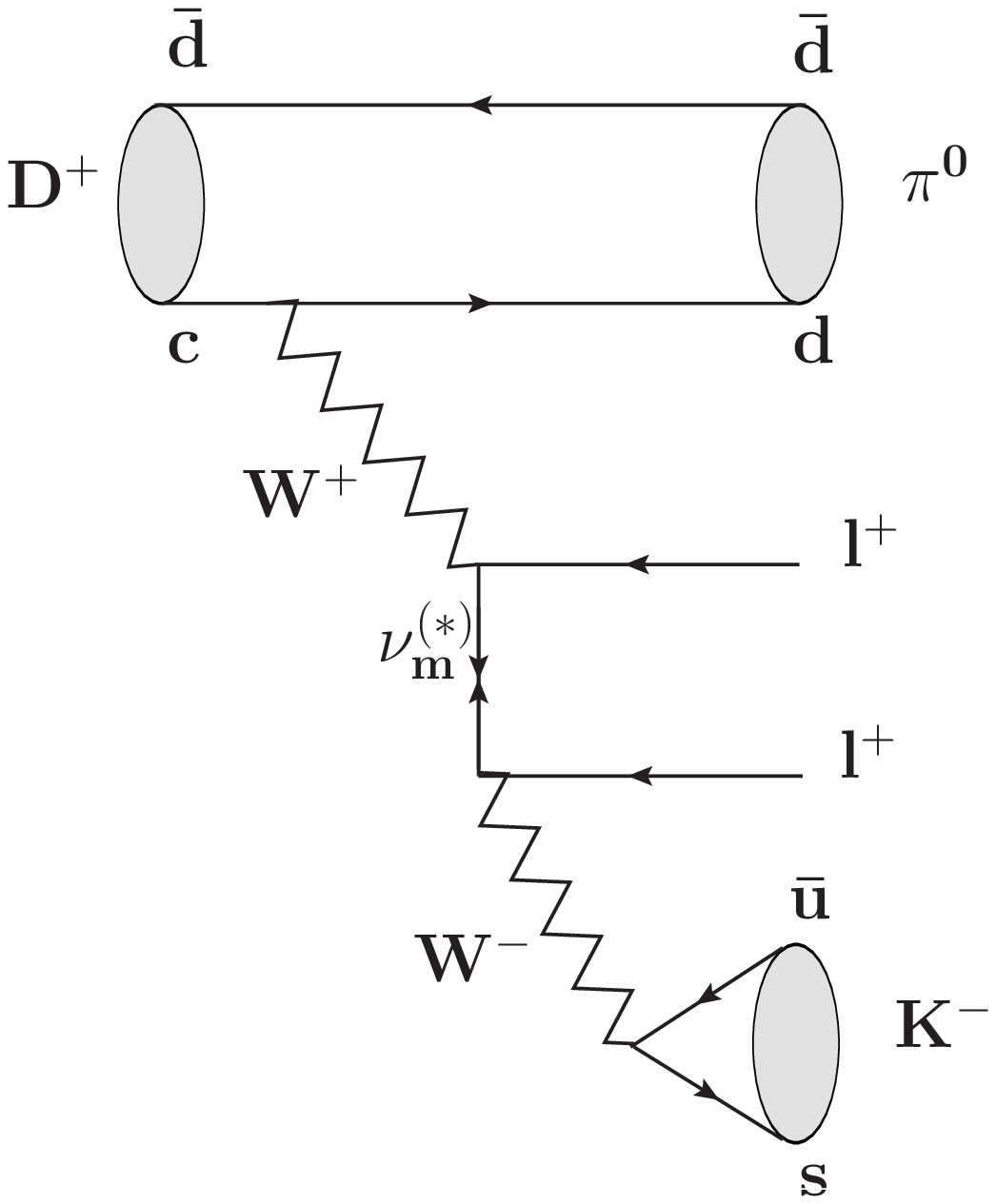}}
\caption{Feynman diagrams for LNV processes $D \to K \pi l^+ l^+$ involving the Majorana neutrino ($\nu_{m}^{(*)}$), where $l$ means the lepton.} \label{chap1:feynmanDiagram}
\end{figure*}

In this paper, we present the studies of LNV processes with $\Delta L$ = 2 in $D$ meson decays $D^0\to K^- \pi^- e^+ e^+$, $D^+ \to K_S^0 \pi^- e^+ e^+$ and $D^+ \to K^- \pi^0 e^+ e^+$. These processes can occur through Cabibbo-favored (CF) and doubly Cabibbo-suppressed (DCS) decays by mediation of a Majorana neutrino, $\nu_m$~\cite{donghairong}, as depicted in Fig.~\ref{chap1:feynmanDiagram}. The DCS processes (Figs.~\ref{fig:subfig:c} and \ref{fig:subfig:d}) are expected to be suppressed by a factor $|V_{cd} V_{us}/V_{cs} V_{ud}| \sim 0.05$~\cite{pdg} with respect to the CF processes (Figs.~\ref{fig:subfig:a} and \ref{fig:subfig:b}). In this analysis, we search for the above three processes as well as the Majorana neutrino with different $m_{\nu_m}$ hypotheses in the CF processes. Additionally, the constraints on the mixing matrix element $|V_{e\nu_m}|^2$ are also estimated depending on $m_{\nu_m}$. The analysis is carried out based on the data sample with an integrated luminosity of 2.93 fb$^{-1}$ at the center-of-mass (C.M.) energy ($\sqrt{s}$) of 3.773 GeV collected with the BESIII detector. Throughout the paper, the charged conjugated modes are always implied implicitly.

\section{\boldmath{Detector and Monte Carlo Simulation}}\label{detectormcsample}
The BESIII detector is a magnetic
spectrometer~\cite{Ablikim:2009aa} located at the Beijing Electron
Positron Collider (BEPCII)~\cite{Yu:IPAC2016-TUYA01}. The
cylindrical core of the BESIII detector consists of a helium-based
 multilayer drift chamber (MDC), a plastic scintillator time-of-flight
system (TOF), and a CsI(Tl) electromagnetic calorimeter (EMC),
which are all enclosed in a superconducting solenoidal magnet
providing a 1.0~T magnetic field. The solenoid is supported by an
octagonal flux-return yoke with resistive plate counter muon
identifier modules interleaved with steel. The acceptance of
charged particles and photons is 93\% over $4\pi$ solid angle. The
charged-particle momentum resolution at $1~{\rm GeV}/c$ is
$0.5\%$, and the $dE/dx$ resolution is $6\%$ for the electrons
from Bhabha scattering. The EMC measures photon energies with a
resolution of $2.5\%$ ($5\%$) at $1$~GeV in the barrel (end cap)
region. The time resolution of the TOF is 68~ps (110~ps) in barrel (end-cap).

Simulated samples produced with the {\sc
geant4}-based~\cite{geant4} Monte Carlo (MC) program which
includes the geometric description of the BESIII detector and the
detector response, are used to determine the detection efficiency
and to estimate the backgrounds. The simulation includes the beam
energy spread and initial state radiation (ISR) in the $e^+e^-$
annihilations modeled with the generator {\sc
kkmc}~\cite{ref:kkmc}.

The cocktail MC sample consists of the production of $D\bar{D}$
pairs with consideration of quantum coherence for all neutral $D$
decay modes, the non-$D\bar{D}$ decays of the $\psi(3770)$, the ISR
production of the $J/\psi$ and $\psi(3686)$ states, and the
continuum processes incorporated in {\sc kkmc}~\cite{ref:kkmc}.
The known decay modes are modeled with {\sc
evtgen}~\cite{ref:evtgen} using BFs taken from the
Particle Data Group~\cite{pdg}, and the remaining unknown decays
from the charmonium states with {\sc
lundcharm}~\cite{ref:lundcharm}. Final state radiation (FSR)
from charged final state particles are incorporated with the {\sc
photos} package~\cite{photos}. The cocktail MC sample is generated to study the possible background sources, and is normalized to the luminosity of the data sample in the analysis.

To study the detector efficiencies of the LNV $\Delta L = 2$ processes, the signal $D$ meson is assumed to decay uniformly in phase space, while in searching for the Majorana neutrino, the exclusive MC samples $D^0 \to K^- e^+ \nu_{m}$ and $D^+ \to K_S^0 e^+ \nu_{m}$ with $\nu_{m}\to \pi^- e^+$ are generated with different $m_{\nu_m}$ assumptions, and the angular distributions are simulated according to the squared amplitude in Eq.~(8) of Ref.~\cite{donghairong}. The form factor is described with the modified pole approximation.

\section{\boldmath{EVENT SELECTION}}
Charged tracks in a candidate event are reconstructed from hits in the MDC.
The charged tracks other than those from $\ks$ decay are required to pass within 10~cm of the interaction point (IP) in the beam direction and within $1$~cm in the plane perpendicular to the beam, as well as satisfy
$|\cos\theta|<0.93$, where $\theta$ is the polar angle relative to the beam direction. The TOF and $dE/dx$ information are combined to determine particle identification (PID)
probabilities ($prob$) for the $\pi$ and $K$ hypotheses, and a $\pi$ ($K$) is identified by requiring $prob(\pi)>prob(K)$ ($prob(K)>prob(\pi)$). To identify an electron or positron, the EMC information is also used to determine the PID probability. An electron or positron is required to satisfy $prob(e)/(prob(e)+prob(\pi) + prob(K)) >0.8$, and $E/p >0.8$, where $E$ and $p$ are the deposited energy in the EMC and the track momentum measured in the MDC, respectively.

The $\ks$ candidates are reconstructed with a vertex-constrained fit for pairs of oppositely charged tracks, assumed to be pions, which are required to pass within $20$ cm of the IP along the beam direction, but with no constraint in the transverse plane. A vertex fit is carried out to insure that the two selected tracks originate from a common vertex, and the fit $\chi^2$ is required to be less than 100. The resulting decay vertex is required to be separated from the IP by greater than twice the resolution. The $\ks$ candidates are further required to have an invariant mass within $[0.487, 0.511]$ GeV/$c^{2}$.

Electromagnetic showers are reconstructed from clusters of energy deposited in the EMC, and the energy
deposited in nearby TOF counters is included to improve the reconstruction efficiency and energy resolution. Photon candidate
showers must have a minimum energy of 25~MeV in the barrel region ($|\cos\theta|<0.80$) or 50~MeV in the end-cap region ($0.86<|\cos\theta|<0.92$).
To suppress showers originating from charged particles, a photon must be separated by at least $10^\circ$ from any charged track. To suppress electronic noise and energy deposits unrelated to the event, timing information from the EMC for the photon candidates must be in coincidence with collision events i.e., $0\leq t \leq 700$ ns. The $\piz$ candidates are reconstructed from pairs of photons. Due to the worse resolution in the EMC end-caps, $\pi^0$ candidates reconstructed with two photons in the end-caps of the EMC are rejected. The invariant mass of two photons is required to be within [0.115,\,0.150] GeV/$c^{2}$ for $\piz$ candidates. In the following analysis, the photon pair is kinematically constrained to the nominal mass of the $\piz$ to improve the resolution of $\piz$ momentum.

In order to improve the positron momentum resolution for the effects of FSR and bremsstrahlung, we use an FSR recovery process, where any photon, which has energy greater than 30 MeV, is separated by more than 20$^{\circ}$ from any shower in the EMC originating from a charged track, and is within a cone of 5$^{\circ}$~around the positron direction, has its momentum added to that of the positron.

In the analysis, the signal candidates of $D$ meson LNV decay are searched for using a single tag (ST) method. Two variables, the beam energy constrained mass $M_{\rm{BC}}$ and the energy difference $\Delta E$,
\begin{eqnarray}
  \label{mbcdeltaE}
   M_{\rm{BC}} &=& \sqrt{E^{2}_{\mathrm{beam}} - |\vec{p}_{D}|^{2}}, \nonumber\\
   \Delta E &=& E_{D} - E_{\mathrm{beam}},
\end{eqnarray}
are used to identify the signal candidates, where $\vec{p}_{D}$ and $E_{D}$ are the momentum and energy of the $D$ candidates in the $e^{+}e^{-}$ C.M. system, and $E_{\mathrm{beam}}$ is the beam energy. The $D$ meson decays form a peak at the nominal $D$ mass in the $M_{\rm{BC}}$ distribution and at zero in the $\Delta E$ spectra. If multiple candidates are present per charm per event, the one with the smallest $|\Delta E|$ is chosen. Candidate events with $M_{\rm{BC}}$ greater than 1.84 GeV/$c^2$ and $\Delta E$ within approximately [-3.5, 2.5] standard deviations of the peak are accepted. The numerical values of the mode dependent $\Delta E$ requirement are listed in Table~\ref{cutdeltae}.

\begin {table}[htbp]
\begin {center}\small
\caption{$\Delta E $ requirements for $D\to K \pi e^+ e^+$ processes.}\label{cutdeltae}
 \setlength{\extrarowheight}{1.0ex}
  \renewcommand{\arraystretch}{1.0}
  \vspace{0.2cm}
\begin {tabular}{p{3.5cm}m{3.8cm}<{\centering}}
\hline \hline
Channel      & $\Delta E$~(MeV)~ \\
\hline
$D^0\to K^- \pi^- e^+ e^+$            & $[-33.0,19.7]$    \\
$D^+ \to K_S^0 \pi^- e^+  e^+$        & $[-30.6, 19.3]$    \\
$D^+ \to K^- \pi^0 e^+ e^+$           & $[-54.8, 24.4]$     \\
\hline
\hline
\end {tabular}
 \vspace{-0.2cm}
\end {center}
\end {table}

Potential background sources are examined with the cocktail MC sample. The dominant contributions are from the processes $\psipp \to D \bar{D}$ with $D\to Ke\nu_{e}$ due to large BFs and the processes $\eeqqbar$, but no peaking background is observed in the $M_{\rm{BC}}$ distribution.

\section{\boldmath Signal determination}\label{fittingresult}
\begin{figure}[hbtp]
  \centering
   \includegraphics[width=0.4\textwidth]{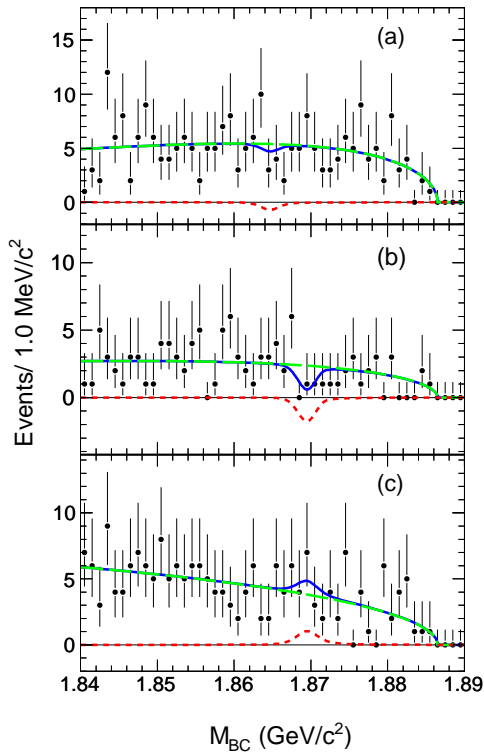}
    \caption{(color online) Fitting on the $M_{\rm{BC}}$ spectra for the decays (a) $D^0\to K^- \pi^- e^+ e^+$, (b) $D^+ \to K_S^0 \pi^- e^+ e^+$ and (c) $D^+ \to K^- \pi^0 e^+ e^+$. The dotted points with error bars are from data, the blue lines are the fitting result, the dashed red and long dashed green lines are the signal and background components, respectively.}
 \label{fitmbccock}
\end{figure}

The signal yields are determined by performing an unbinned maximum likelihood fit on the $M_{\rm BC}$ distribution of surviving candidate events. In the fit, the background shape is described by an ARGUS function~\cite{argus}, and the signal shape is modeled by the MC simulated shape convolved with a Gaussian function which accounts for the resolution difference between data and MC simulation. The width of the Gaussian function is fixed to be 0.32 MeV/$c^{2}$, obtained from a control sample of $\dkpipipi$ decay. The fits are shown in Fig.~\ref{fitmbccock}. The BFs, $\mathcal{B}_{\dhhee}$, are calculated by
\begin{eqnarray}
  \label{calbranch}
  \mathcal{B}_{D\to K \pi e^+ e^+} = \frac{N_{\rm sig}}{2 \cdot N^{\rm tot}_{\rm D \bar{D}} \cdot \epsilon \cdot \mathcal{B}},
\end{eqnarray}
where $N_{\rm{sig}}$ is the signal yield determined from the fit, $N_{D\bar{D}}^{\rm{tot}}$ is the total number of $D \bar{D}$ pairs, which are $(8,296\pm 31\pm 65)\times 10^{3}$ for $D^+ D^-$ pairs and $(10,597 \pm 28 \pm 98) \times 10^3$ for $D^0 \bar{D}^0$ pairs~\cite{crosssection}, $\epsilon$ is the detection efficiency, obtained from the corresponding MC simulation, and $\mathcal{B}$ is the decay branching fraction of the intermediate state, i.e., 1 in the decay $D^0 \to K^- \pi^- e^+ e^+$ due to no intermediate state involved, $\mathcal{B}_{K_S^0 \to \pi^+ \pi^-}$ in the decay $D^+ \to K^0_{S} \pi^- e^+ e^+$ and $\mathcal{B}_{\pi^0 \to \gamma \gamma}$ in the decay $D^+ \to K^- \pi^0 e^+ e^+$, where $\mathcal{B}_{K_S^0 \to \pi^+ \pi^-}$ and $\mathcal{B}_{\pi^0 \to \gamma \gamma}$ are taken from the world average values~\cite{pdg}. A factor of 2 in the denominator indicates both $D$ and $\bar{D}$ mesons in every event are included.

Since no obvious signal is observed, the ULs at the 90\% confidence level~(CL) on the BFs of $\dhhee$ decays are set after considering the effect of systematic uncertainties.

\section{\boldmath SYSTEMATIC UNCERTAINTIES}\label{Sys_err}
The systematic uncertainties arise from several aspects including the tracking and PID efficiencies of charged tracks, $\ks$ and $\piz$ reconstruction efficiencies, total number of $D\bar{D}$ pairs, BFs of $K_{S}^{0} \to \pi^{+} \pi^{-}$ and $\pi^{0} \to \gamma \gamma$ decays, $\Delta E$ requirement, FSR recovery, modeling for detection efficiency and fitting $M_{\rm{BC}}$.

Systematic uncertainties from the tracking efficiency of $K$, $\pi$ and $e$ are assigned to be 1.0\% per track~\cite{fangy_systrack, qinxiaoshuaiomegapi}. For the PID efficiency, the systematic uncertainties for $K(\pi)$ and $e$ are 0.5\% and 1.0\% per track~\cite{fangy_systrack, qinxiaoshuaiomegapi}, respectively. Systematic uncertainties from $\ks$ and $\piz$ reconstruction are taken to be 1.5\% and 2\%~\cite{ksmunu_huangzl}, respectively.

The systematic uncertainty of the total number of $D\bar{D}$ pairs is 0.9\%~\cite{crosssection}. The BFs of $K_{S}^{0} \to \pi^{+} \pi^{-}$ and $\pi^{0} \to \gamma \gamma$ are ($69.20\pm 0.05$)\% and ($98.823\pm 0.034$)\% from the world average values~\cite{pdg}, resulting in the systematic uncertainty of 0.1\% and 0.0\%, respectively.

The systematic uncertainty from the $\Delta E$ requirement is studied using control samples of $\dkpipizero$ and $\dkpipi$ for the signal processes with and without $\pi^0$ in final states, where the control samples are selected with the ST method. We set [$\mu-3.5\sigma$, $\mu+2.5\sigma$] as a nominal $\Delta E$ window for the signal, where $\mu$ and $\sigma$ are the mean and width values of $\Delta E$ obtained by fitting. Then we vary the $\Delta E$ window by 0.5$\sigma$ on both sides, and the resulting differences of the change of efficiency between data and MC simulation are taken as the systematic uncertainties.

To study the systematic uncertainty associated with FSR recovery process, we obtain the alternative detection efficiency without the FSR recovery process, and the difference in the efficiency is taken as the systematic uncertainty.

The difference of the geometric efficiency between the one obtained with the phase space generator, and the average of $m_{\nu_m}$-dependent cases, is taken as the systematic uncertainty associated with the modeling.

The systematic uncertainty associated with the fitting of the $M_{\rm{BC}}$ distribution arises from the fitting range, signal shape and background shape. We performed alternative fits by varying the fitting range from [1.84, 1.89] to [1.845, 1.89] GeV/$c^{2}$, the width of convolved Gaussian for signal shape within one standard deviation, and the background shape from the ARGUS function to the cocktail MC simulated shape. The relative changes of the signal yields are taken as the corresponding systematic uncertainties, and are found to be negligible compared to the statistical uncertainties.

All the systematic uncertainties are summarized in Table~\ref{chap7:tableSysSummary}. Assuming they are independent, the total systematic uncertainty is the quadrature sum of the individual ones.

\begin {table*}[htbp]
\begin {center}\small
\caption{Relative systematic uncertainties for the $\dhhee$ processes (in percent). Here `--' denotes that there is no corresponding systematic uncertainty, and `...' means that the corresponding systematic uncertainty is negligible.}\label{chap7:tableSysSummary}
\setlength{\extrarowheight}{1.0ex}
  \renewcommand{\arraystretch}{1.0}
  \vspace{0.2cm}
\begin {tabular}{l|c|c|c}
\hline \hline
\multirow{2}{*}{Source}& \multicolumn{3}{c}{Relative systematic uncertainty (\%)} \\
\cline{2-4}
                   & $D^0\to K^- \pi^- e^+  e^+$ & $D^+ \to K_S^0 \pi^- e^+ e^+$ & $D^+ \to K^- \pi^0 e^+  e^+$ \\
\hline
Tracking                   & 4.0      & 3.0      & 3.0  \\
PID                        & 3.0      & 2.5      & 2.5   \\
$K_{S}^{0}$ selection      & --       & 1.5      & --   \\
$\pi^{0}$ selection        & --       & --       & 2.0    \\
$N_{D\bar{D}}$             & 1.0      & 0.9   &0.9     \\
Cited BF                   &--        & 0.1   &0.0     \\
$\Delta E$ requirement     & 0.7      & 0.7   &0.4     \\
FSR recovery               & 0.6      & 0.8   &0.6     \\
Efficiency modeling        & 3.6      & 4.3   &4.7      \\
Fitting $M_{\rm{BC}}$      & ...      & ...   & ...\\
\hline
Total                      &6.3       & 6.2    & 6.5 \\
\hline
\hline
\end {tabular}
\vspace{-0.2cm}
\end {center}

\end {table*}

\section{\boldmath Results and discussion }

\subsection{\boldmath Upper limits for $\dhhee$ decays }

Taking into account the effect of systematic uncertainties, we calculate ULs on the BFs for the LNV $\Delta L = 2$ decays $D^0\to K^- \pi^- e^+  e^+$, $D^+ \to K_S^0 \pi^- e^+  e^+$ and $D^+ \to K^- \pi^0 e^+ e^+$ according to Eq.~(\ref{calbranch}) based on the Bayesian method~\cite{bayesian}. A series of fits of the $M_{\rm{BC}}$ distribution are carried out fixing the BF at different values, and the resultant curve of likelihood values as a function of the BF is convolved with a Gaussian function, which has a width given by the overall systematic uncertainty and is normalized to the maximum value of 1. The ULs on the BF at the 90\% CL, $\mathcal{B}^{\rm UL}_{\rm sig}$ for the different processes, which are listed in Table~\ref{chap3:upperlimitMCResults}, are the values that yield 90\% of the likelihood integral over BF from zero to infinity.

\begin {table}[hbtp]
\begin {center}
\caption{The detection efficiencies ($\epsilon$), the ULs at the 90\% CL on the signal yields ($N^{\rm{UL}}_{\rm{sig}}$), and the BFs ($\mathcal{B}^{\rm{UL}}_{\rm{sig}}$) of $\dhhee$ processes.}\label{chap3:upperlimitMCResults}
\setlength{\extrarowheight}{1.0ex}
  \renewcommand{\arraystretch}{1.0}
  \vspace{0.2cm}
\begin {tabular}{p{3.5cm}m{1.2cm}<{\centering}m{1.2cm}<{\centering}m{1.5cm}<{\centering}}
\hline\hline
Channel &$\epsilon(\%)$& $N^{\mathrm{UL}}_{\rm{sig}}$ & $\mathcal{B}^{\mathrm{UL}}_{\rm{sig}} (\times10^{-6})$ \\
\hline
$D^0\to K^- \pi^- e^+  e^+$    &16.8 &10.0  & $<2.8$ \\
$D^+ \to K_S^0  \pi^- e^+ e^+$ &11.5 &4.4  & $<3.3$ \\
$D^+ \to K^- \pi^0 e^+  e^+$   &10.6 &14.8 & $<8.5$ \\
\hline
\hline
\end {tabular}
\vspace{-0.2cm}
\end {center}
\end {table}

\subsection{\boldmath Searching for Majorana neutrino}
\label{resultandiscussion}

With the above three decay processes, the Majorana neutrino can be searched for by studying the decay chains $D^0 \to K^- e^+ \nu_{m}(\pi^- e^+)$, $D^+ \to K_S^0 e^+ \nu_{m}(\pi^- e^+)$ or $D^+ \to \pi^0 e^+ \nu_{m}(K^- e^+)$; a narrow peak will be present in the distribution of $\pi^- e^+$ ($K^- e^+$) invariant mass if a signal exists. Compared to the other two decay channels, the $D^+ \to \pi^0 e^+ \nu_{m}(K^- e^+)$ is expected to be suppressed by a factor of $1/20$ because of the smaller CKM factors. So in this analysis, the Majorana neutrino is searched in the processes $D^0 \to K^- e^+ \nu_{m}(\pi^- e^+)$ and $D^+ \to K_S^0 e^+ \nu_{m}(\pi^- e^+)$ with different $m_{\nu_m}$ hypotheses, i.e., from 0.25 to 1.0 GeV/$c^{2}$ with an interval of 0.05 GeV/$c^2$.

\begin{figure}[htbp]
\begin{center}
 \includegraphics[width=0.4\textwidth, height=8.1cm]{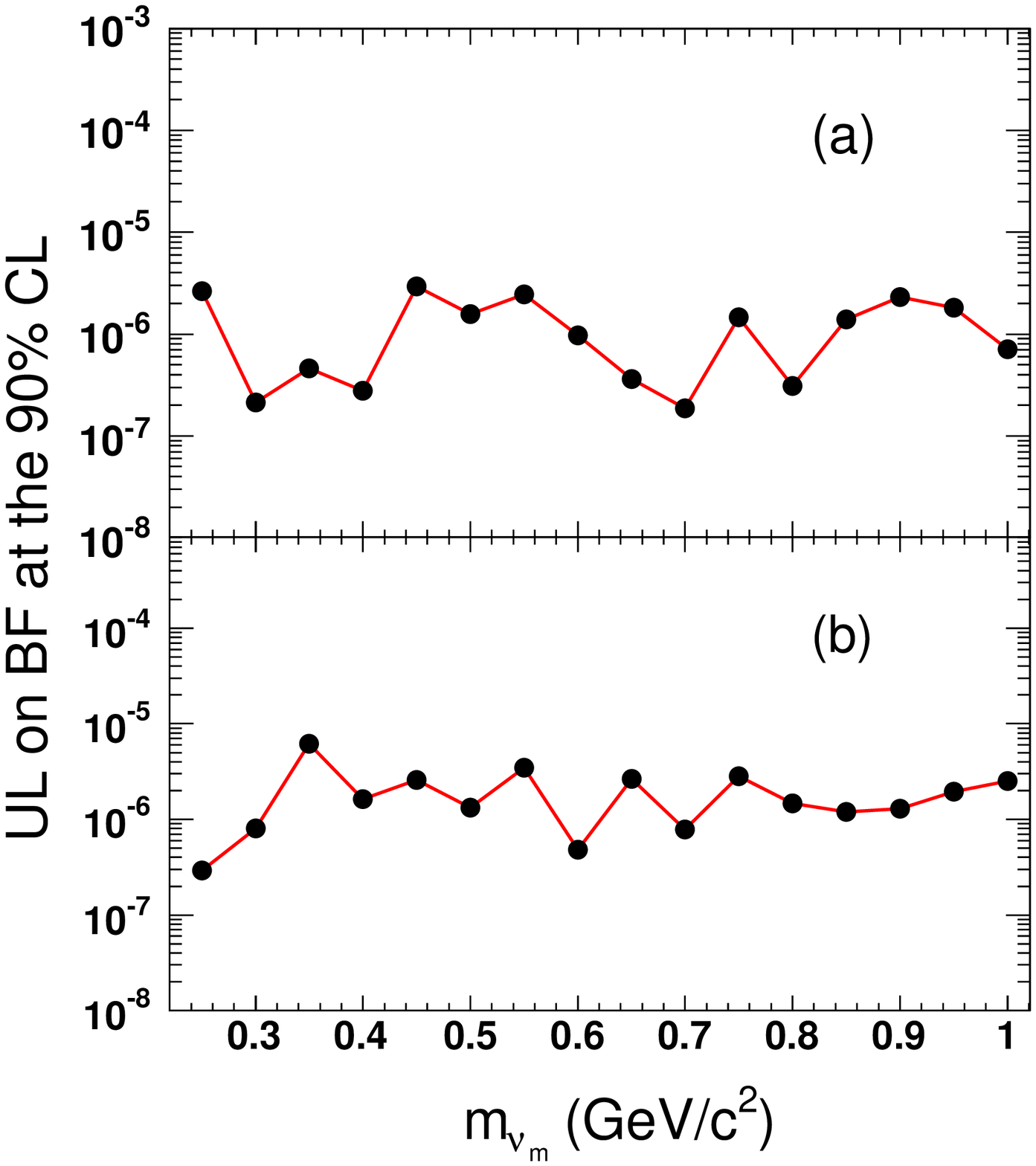}
 \includegraphics[width=0.4\textwidth, height=8.1cm]{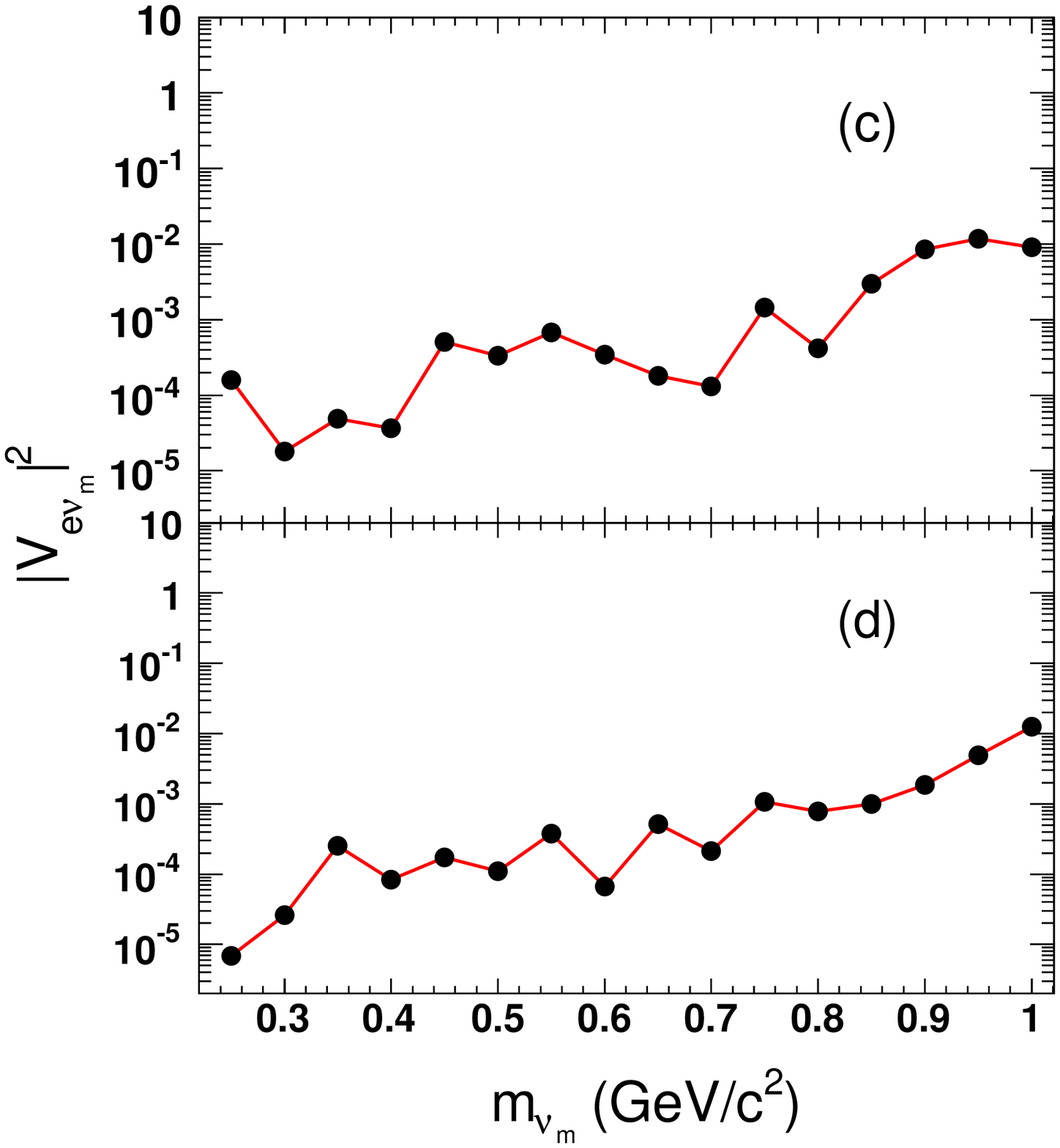}
  \caption{The ULs on the (a)(b) BF and the (c)(d) mixing matrix element $|V_{e\nu_m}|^{2}$ at the 90\% CL as a function of $m_{\nu_m}$ for the decays (a)(c) $D^0\to K^- e^+ \nu_m (\pi^- e^+)$ and (b)(d) $D^+ \to K_S^0 e^+ \nu_m (\pi^- e^+)$.}
  \label{chap3:massRelatedUpperLimit}
\end{center}
\end{figure}

Based on the above selection criteria, to search for the Majorana neutrino with a given mass, $m_{\nu_m}$, the candidate events are selected by further requiring the invariant mass of any $\pi^- e^+$ combination (two $e^+$ per event), $M_{\pi^- e^+}$, to be within the range of [$m_{\nu_m}-3\sigma$, $m_{\nu_m}+3\sigma$], where $\sigma$ is the resolution of the $M_{\pi^- e^+}$ distribution obtained by studying the signal MC sample. The number counting method is used to determine the signal yields due to very few events surviving. We count the number of signal candidates within the $M_{\rm{BC}}$ signal region of [1.859, 1.872] ([1.865, 1.875]) GeV/$c^{2}$ for the decay $D^0\to K^- e^+ \nu_m (\pi^- e^+)$ ($D^+ \to K_S^0 e^+ \nu_m (\pi^- e^+)$). The number of background events is estimated from the side-band regions of the $M_{\rm BC}$ distribution, defined as [1.842, 1.852] and [1.876, 1.886] ([1.842, 1.854] and [1.878, 1.886]) GeV/$c^{2}$, taking into account the scale factor obtained by fitting the $M_{\rm BC}$ distribution as shown in Fig.~\ref{fitmbccock}. The ULs on the BFs of Majorana neutrino case are calculated with the profile likelihood method incorporating the systematic uncertainty with {\sc trolke}~\cite{trolke1,trolke2} in the {\sc root} framework, where the numbers of events in the signal and side-band regions are assumed to be described by Poisson distributions and the efficiency by a Gaussian distribution. The ULs on the BFs at the 90\% CL as a function of $m_{\nu_m}$ are at the level of $10^{-7}\sim10^{-6}$, as shown in Fig.~\ref{chap3:massRelatedUpperLimit}.

Based on the measured BFs, the mixing matrix element $|V_{e\nu_m}|^{2}$ of a positron with the heavy Majorana neutrino in the charged current interaction~\cite{feynman1,uppersummary} as a function of $m_{\nu_m}$ can be obtained by Eq.~(\ref{calven})~\cite{donghairong},

\begin{eqnarray}
  \label{calven}
   \frac{\Gamma(m_{\nu_m},V_{e\nu_m}(m_{\nu_m}))}{\Gamma(m_{\nu_m},V^{'}_{e\nu_m}(m_{\nu_m}))} = \frac{|V_{e\nu_m}(m_{\nu_m})|^{4}}{|V^{'}_{e\nu_m}(m_{\nu_m})|^{4}},
\end{eqnarray}
where the decay width $\Gamma(m_{\nu_m},V_{e\nu_m}(m_{\nu_m}))$ is proportional to its BF, and $\Gamma(m_{\nu_m},V^{'}_{e\nu_m}(m_{\nu_m}))$ is related to the BF given in Tables~4 and 5 of Ref.~\cite{donghairong}, based on the assumptions that the Majorana neutrino is on-shell and its width is negligible compared to the neutrino mass. The mixing matrix element $|V^{'}_{e\nu_m}(m_{\nu_m})|^{2}$ is derived from a reanalysis of neutrinoless double beta decay experimental data~\cite{doublebeta}. The resultant ULs on the mixing matrix element $|V_{e\nu_m}|^{2}$ as a function of $m_{\nu_m}$, which are also depicted in Fig.~\ref{chap3:massRelatedUpperLimit}, provide complementary information in $D$ meson decays.

\section{\boldmath SUMMARY}\label{summary}

Using the data sample with the integrated luminosity of $2.93~\mathrm{fb}^{-1}$ collected at the C.M. energy $\sqrt{s} = 3.773$ GeV, we perform a search for LNV $\Delta L =2$ decays of $\dhhee$ as well as search for a Majorana neutrino with different mass hypotheses. No evidence of a signal is found. Therefore, using the Bayesian approach, we place 90\% CL ULs on the decay BFs for $D^0\to K^- \pi^- e^+  e^+$, $D^+ \to K_S^0  \pi^- e^+ e^+$ and $D^+ \to K^- \pi^0 e^+ e^+$ to be $2.8\times10^{-6}$, $3.3\times10^{-6}$ and $8.5\times10^{-6}$, respectively. We also determine ULs, which are of the level $10^{-7}\sim10^{-6}$, on the BFs at the 90\% CL for the decays $D^0\to K^- e^+ \nu_m (\pi^- e^+)$ and $D^+ \to K_S^0 e^+ \nu_m (\pi^- e^+)$ with different $m_{\nu_m}$ hypotheses within the range 0.25 to 1.0 GeV/$c^2$. The constraints on the mixing element $|V_{e\nu_m}|^{2}$ depending on $m_{\nu_m}$ are also evaluated based on the related variables from Ref.~\cite{donghairong} and the measured BFs. The results provide the supplementary information in the study of mixing between the heavy Majorana neutrino and the standard model neutrino $\nu_e$ in $D$ meson decays.

\vspace{8pt}
\section*{\boldmath ACKNOWLEDGEMENTS}
The BESIII collaboration thanks the staff of BEPCII and the IHEP computing center for their strong support. This work is supported in part by National Key Basic Research Program of China under Contract No. 2015CB856700; National Natural Science Foundation of China (NSFC) under Contracts Nos. 11805037, 11235011, 11335008, 11425524, 11625523, 11635010; the Chinese Academy of Sciences (CAS) Large-Scale Scientific Facility Program; the CAS Center for Excellence in Particle Physics (CCEPP); Joint Large-Scale Scientific Facility Funds of the NSFC and CAS under Contracts Nos. U1832121, U1332201, U1532257, U1532258; CAS Key Research Program of Frontier Sciences under Contracts Nos. QYZDJ-SSW-SLH003, QYZDJ-SSW-SLH040; 100 Talents Program of CAS; National 1000 Talents Program of China; INPAC and Shanghai Key Laboratory for Particle Physics and Cosmology; German Research Foundation DFG under Contracts Nos. Collaborative Research Center CRC 1044, FOR 2359; Istituto Nazionale di Fisica Nucleare, Italy; Koninklijke Nederlandse Akademie van Wetenschappen (KNAW) under Contract No. 530-4CDP03; Ministry of Development of Turkey under Contract No. DPT2006K-120470; National Natural Science Foundation of China (NSFC) under Contracts Nos. 11505034, 11575077; National Science and Technology fund; The Swedish Research Council; U. S. Department of Energy under Contracts Nos. DE-FG02-05ER41374, DE-SC-0010118, DE-SC-0012069; University of Groningen (RuG) and the Helmholtzzentrum fuer Schwerionenforschung GmbH (GSI), Darmstadt; WCU Program of National Research Foundation of Korea under Contract No. R32-2008-000-10155-0.


\end{document}